\documentclass[twocolumn,secnumarabic,amssymb, nobibnotes, aps, prl, superscriptaddress]{revtex4-2}

\usepackage[]{graphicx}
\usepackage{tabularx}
\usepackage[usenames,dvipsnames]{color}
\usepackage{soul}
\usepackage{ulem}
\usepackage{bm}
\usepackage{amsmath}
\usepackage{lineno}
\usepackage{gensymb}

\usepackage{times}
\usepackage{amsfonts}
\usepackage{mathrsfs}
\usepackage{graphicx}
\usepackage{dcolumn}
\usepackage{bm}
\usepackage{color}

\usepackage[colorlinks,bookmarks=false,citecolor=blue,linkcolor=red,urlcolor=blue]{hyperref}
\usepackage{multirow}
\usepackage{physics}

\begin{document}
\title{Back-action supercurrent rectifiers} 
\author{Daniel Margineda}
\email{d.margineda@nanogune.eu, present address: CIC nanoGUNE BRTA, E-20018 Donostia-San Sebastián, Spain.}
\affiliation{NEST, Istituto Nanoscienze-CNR and Scuola Normale Superiore, I-56127, Pisa, Italy}

\author{Alessandro Crippa}
\affiliation{NEST, Istituto Nanoscienze-CNR and Scuola Normale Superiore, I-56127, Pisa, Italy}
\author{Elia Strambini}
\affiliation{NEST, Istituto Nanoscienze-CNR and Scuola Normale Superiore, I-56127, Pisa, Italy}
\author{Laura  Borgongino}
\affiliation{NEST, Istituto Nanoscienze-CNR and Scuola Normale Superiore, I-56127, Pisa, Italy}

\author{Alessandro Paghi}
\affiliation{NEST, Istituto Nanoscienze-CNR and Scuola Normale Superiore, I-56127, Pisa, Italy}
\author{Giorgio de Simoni}
\affiliation{NEST, Istituto Nanoscienze-CNR and Scuola Normale Superiore, I-56127, Pisa, Italy}
\author{Lucia Sorba}
\affiliation{NEST, Istituto Nanoscienze-CNR and Scuola Normale Superiore, I-56127, Pisa, Italy}
\author{Yuri Fukaya}
\affiliation{SPIN-CNR, I-84084 Fisciano (SA), Italy}
\author{Maria Teresa Mercaldo}
\affiliation{Dipartimento di Fisica ``E. R. Caianiello", Universit\`a di Salerno, I-84084 Fisciano (SA), Italy}
\author{Carmine Ortix}
\affiliation{Dipartimento di Fisica ``E. R. Caianiello", Universit\`a di Salerno, I-84084 Fisciano (SA), Italy}
\author{Mario Cuoco}
\affiliation{SPIN-CNR, I-84084 Fisciano (SA), Italy}
\author{Francesco Giazotto}
\email{francesco.giazotto@sns.it}
\affiliation{NEST, Istituto Nanoscienze-CNR and Scuola Normale Superiore, I-56127, Pisa, Italy}

\begin{abstract}
Back-action refers to a response that retro-acts on a system to tailor its properties with respect to an external stimulus. This effect is at the heart of many electronic devices such as amplifiers, oscillators, and sensors.  Here, we demonstrate that back-action can be exploited to achieve \textit{non-reciprocal}  transport in superconducting circuits. 
In our devices, dissipationless current flows in one direction whereas dissipative transport occurs in the opposite direction. Supercurrent diodes presented so far rely on magnetic elements or vortices to mediate charge transport or external magnetic fields to break time-reversal symmetry.  Back-action solely turns a conventional reciprocal superconducting weak link with no asymmetry between the current bias directions into a rectifier, where the critical current amplitude depends on the bias sign. The self-interaction of the supercurrent stems from the gate tunability of the critical current in metallic and semiconducting systems, which promotes nearly ideal magnetic field-free rectification with selectable polarity. 

\end{abstract}
\maketitle
\noindent\large{\textbf{Introduction}}\normalsize\\ 
Control of dissipationless transport is a core challenge for superconducting electronics and it is at the heart of several applications including both classical~\cite{lik91} and quantum~\cite{cla08} computation. 
The flow of dissipationless current relies on macroscopic quantum coherence and exploits the phase difference between spatially separated superconducting condensates. The resulting supercurrent cannot exceed a critical amplitude, $I_c$, above which the superconductor turns into the normal state.
In analogy with the semiconducting counterpart, the superconducting diode effect 
refers to dissipationless current flowing in one direction $(+)$ while the current is driven by dissipative carriers in the opposite direction $(-)$, thereby leading to asymmetric amplitudes of the critical currents $I_c^+ \neq |I_c^-|$.

The discovery and design of quantum materials and platforms suitable for nonreciprocal superconducting transport have been developed around the fundamental requirement of realizing the lack of inversion and time-reversal symmetries~\cite{and20,wu22,jeo22,bau22,baur22,pal22}. 
This condition is, for instance, realized in systems that are naturally equipped with symmetry-breaking crystalline potentials and magnetic interactions, as with noncentrosymmetric or magnetic materials~\cite{and20, wu22, tra23, gol22, gut23, Narita2022,tra23,lin22} and related heterostructures~\cite{nad23}. 
Alternatively, several implementations exploit external magnetic fields~\cite{Edelstein95,wak17,and20,Yuan22,Lyu2021} leading to symmetry-breaking configurations, which yield nonreciprocal superconducting transport. 
Hence, the current paradigm for superconducting diodes either relies on internal mechanisms~\cite{dai22,he22,Ilic22,Scammell_2022}, 
or on suitably designed external sources to break time and spatial symmetry~\cite{sur22,hou23,sun23,fuk24}. 

 We propose a general mechanism to achieve nonreciprocal superconducting transport that arises from the back-action of the supercurrent on a reciprocal weak link. Differently from self-field back-action mechanisms~\cite{kras97,gol22}, our scheme for supercurrent rectification is magnetic field-free.
 The idea of exploiting a back-action mechanism not linked to internal or external sources of symmetry breaking aims at a general design principle suitable for several superconducting platforms.
\begin{figure*}[ht]
\centering
\includegraphics[scale=0.45 ]{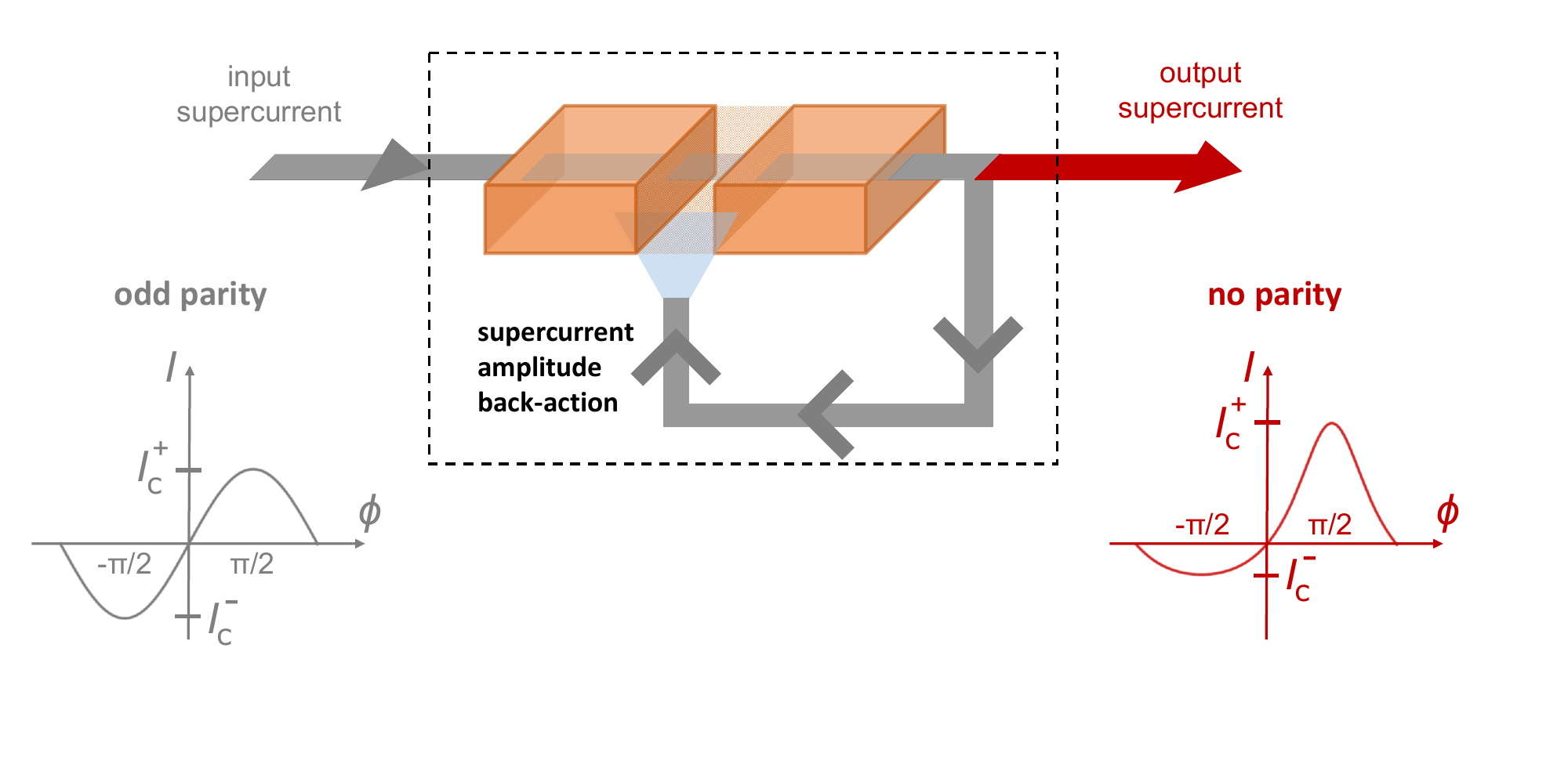}
\caption{
\textbf{Principle of supercurrent rectifier by supercurrent amplitude back-action.} Cartoon of a generic weak link with two superconducting banks (orange) separated by an element with weaker superconductivity (orange with texture). A period of the current-phase relation of the input current injected from the left lead is shown, and it is odd-in-$\phi$. The supercurrent coming out from the right lead retroacts on the weak link itself, thereby generating an output supercurrent with no defined parity on $\phi$. The odd parity of the input current-phase relation dictates $I_c^+=|I_c^-|$, while the back-action results in an output supercurrent without defined parity in $\phi$, so that $I_c^+\neq|I_c^-|$. See the main text for a model of the back-action mechanism.}
\label{implementation} 
\end{figure*}
We demonstrate how to achieve a back-action on the supercurrent amplitude via an applied gate voltage to a superconducting weak link. 
This self-induced effect provides a fundamental path to rectify the supercurrent, thereby leading to a tunable and high-rectification efficiency.
The back-action of the supercurrent on the effective gate voltage relies on a control resistor in series with the Josephson element. The gate voltage modifies the supercurrent, which in turn alters the effective gate voltage experienced by the weak link, thereby realizing a retroaction. 
Our findings set out a general paradigm for the design of all-electrical and magnetic field-free supercurrent rectifiers.\\

\noindent\large{\textbf{Results}}\normalsize\\
\textbf{Device concept.}
The magnitude of the dissipationless supercurrent $I$ through a Josephson element is entangled to the superconducting phase difference $\phi$ across the junction via the so-called current-phase relation $I(\phi)$. 
In its simplest form, it reads $I(\phi)=I_c \sin{(\phi)}$, and in generic reciprocal junctions it is an odd $2\pi$-periodic function of $\phi$.
From the functional form above, it follows immediately the reciprocity of the two critical currents, as depicted in the bottom left sketch of Fig.~\ref{implementation}: $I_c^+=\max (I(\phi))= |I_c^-|=|\min (I(\phi))|=I_c$.
The reciprocity can be violated in systems that concomitantly break inversion and time-reversal symmetry either intrinsically or using applied fields. 
The concept of a back-action supercurrent rectifier instead relies on a modulation of $I_c$ induced by the flowing current $I$. 
For the sake of simplicity, we assume a linear modulation of the form $I_c=I_c^0+\alpha I$, where $-1<\alpha<1$ represents the back-action strength, and $I_c^0$ is the critical current in the absence of back-action ($\alpha=0$). 
Other monotonic functions $I_c(I)$ bring similar conclusions, provided that $I_c$ and $\phi$ are factorized in the current-phase relation, i.e., $I(\phi)=f(I)g(\phi)$.  
Solving the self-consistent equation $I=(I_c^0+\alpha I)\sin{\phi}$ yields $I(\phi)=I_c^0\frac{ \sin{\phi} }{1-\alpha \sin{\phi} }$, which is a functional form with a non-defined parity, as sketched in the bottom right of  Fig~\ref{implementation}.
It is straightforward to show that the above function is non-reciprocal, with $|I_c^{\pm}|=\frac{I_c^0}{1\mp\alpha}$ and rectification efficiency  $\eta \equiv (I_c^+-|I_c^{-}|)/(I_c^++|I_c^{-}|)=\alpha$. Hence, in the limit of $|\alpha| \simeq 1$ an ideal rectification can be achieved.
The model so far described, though very minimal, captures the essential features of back-action: i) parity violation of the current-phase relation, ii) zero spontaneous supercurrent [i.e., $I(0)=0$], and iii) nonreciprocal supercurrent and tunable rectification amplitude (See Supplementary Note for a formal discussion).\\

\begin{figure*}[ht]
\centering
\includegraphics[scale=0.2 ]{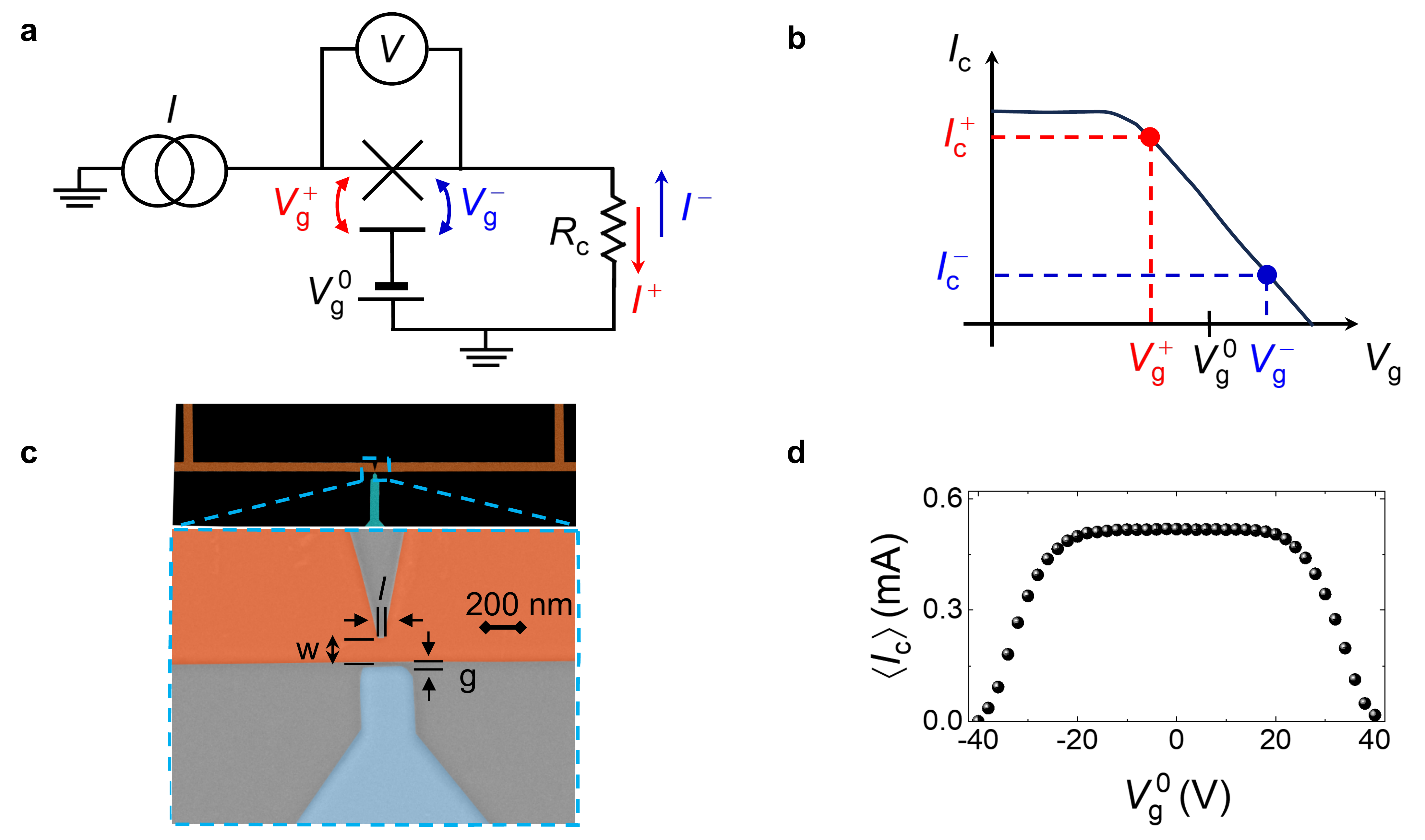}
\caption{\textbf{Setup for the back-action supercurrent rectifier.} \textbf{a,} Circuit schematic to realize a supercurrent rectifier by back-action mechanism. The cross represents a weak link measured in a current-biased ($I$) setup by monitoring the voltage drop across it ($V$). A gate electrode at nominal voltage $V_g^0$ tunes the critical current of the weak link nearby. A control resistor in series, $R_c$, modifies $V_g^0$ into $V_g=V_g^0-I R_c$, depending on the sign of the current bias $I$. \textbf{b,} Effect of the circuitry presented in \textbf{a} on a representative critical current vs. gate voltage trace. At a certain gate voltage bias $V_g^0$, the positive critical current $I_c^+$ (red dot) at $V_g^+$ differs from the negative critical current $I_c^-$ (blue dot) at $V_g^-$ resulting in a finite diode effect, i.e., $I_c^+ \neq |I_c^-|$. \textbf{c}, Scanning electron micrograph of a niobium (Nb) strip (orange) with a nanoconstriction (characteristic dimensions $l\sim 80$\,nm and $w\sim180$\,nm) to implement the superconducting weak link. The gate voltage $V_g^0$ is applied through a side gate (light blue) located at distance $g\sim 50$\,nm.
\textbf{d}, Experimental gate voltage dependence of the average critical current $\langle I_c \rangle = (I_{c}^+ +|I_{c}^-|)/2$ recorded at $T=1$\,K with $R_c=$2\,k$\Omega$.}
\label{implementation1} 
\end{figure*}

\noindent\textbf{Implementation of a back-action supercurrent rectifier.}
We now turn to the physical realization of the back-action mechanism. First, we introduce the circuitry that allows the operation of a generic \textit{gate-tunable} Josephson junction as a highly efficient supercurrent rectifier. Later, we present the sample of the experiment and characterize its performance.

Figure~\ref{implementation1}a shows the schematic of the current-biased setup. The cross represents a weak link whose critical current is measured by sweeping the current bias $I$ till the transition to the normal state is detected by a finite voltage drop $V$ across the weak link. Red and blue bias currents correspond to positive and negative sweep directions, respectively.  A third control terminal tunes the critical current via a voltage source at $V_g^0$. 
The feedback network consists of a control resistor $R_c$ embedded between the weak link and the ground. 
Thus, the bias current $I$ lifts upward or downward the weak link potential according to its direction resulting in a polarity-dependent gate voltage, $V_g=V_g^0-I R_c$.

Figure~\ref{implementation1}b illustrates the superconducting rectifier principle: 
at a voltage bias $V_g^0$ for which $I_c$ depends on the gate voltage 
(i.e., $\partial I_c / \partial V_g \ne 0$),
the critical current for a positive bias current $I_c^+$ (red) differs in modulus from the critical current for a negative bias current $|I_c^-|$ (blue), thereby resulting in $I_c^+\neq |I_c^-|$. 
As shown below, the strength of the non-reciprocity depends both on the magnitude of $R_c$ and on  $\partial I_c / \partial V_g$, which we shall refer to as transconductance $g_m$.
It is worth mentioning that, in general, back-action rectifiers can be designed to operate without a voltage source on the gate, i.e., at $V_g^0=0$. In non-symmetric systems ($I_c(V_g^0) \neq I_c(-V_g^0)$), the supercurrent polarity-dependent gate voltage $IR_c$ promotes rectification at any $V_g^0$, making the battery on the gate unnecessary.

\begin{figure*}[ht]
\centering
\includegraphics[scale=0.16 ]{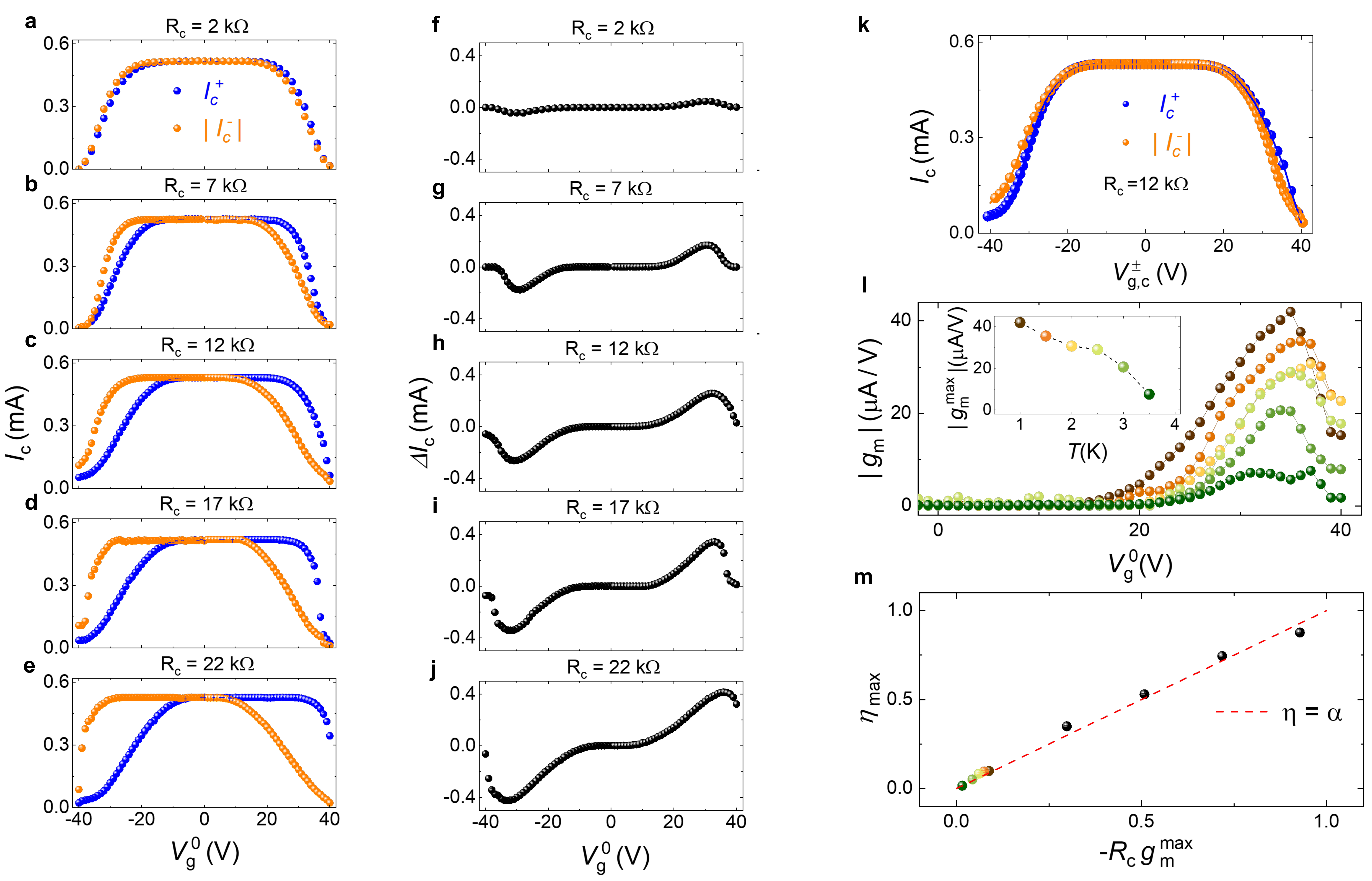}
\caption{
\textbf{Back-action supercurrent rectifier  characterization.} \textbf{a-e,} Gate voltage dependence of the switching current for positive $I_{c}^+$ (blue dots) and negative $|I_{c}^-|$(orange dots) bias current for different values of the control resistor $R_c$. The diode effect results in $I_{c}^+\neq |I_{c}^-|$ and is enhanced by the increase of $R_c$ from 2\,k$\Omega$  to 22\,k$\Omega$.
\textbf{f-j,} $\Delta I_{c}= I_{c}^+-|I_{c}^-|$ as a function of $V_g^0$ obtained from the data in \textbf{a-e}. \textbf{k}, Quasi-symmetric behavior the critical currents obtained by normalizing the gate voltage with the back-action contribution, $V_{g,c}^{\pm}=V_g^0 -I_{c}^{\pm}R_c$ for $R_c= 12$\,k$\Omega$. Similar plots for all the configurations (variation of $R_c$ or temperature) are in Supplementary Figure\,2. \textbf{j,} Transconductance $g_m=\partial \langle I_c \rangle / \partial V_g^0$ as a function of gate voltage $V_g^0$ for different bath temperatures $T$. Color code is given in the inset where the maximum transconductance $g_m^{max}$ versus temperature $T$ is plotted. \textbf{m} Rectification efficiency $\eta$ evaluated at its maximum value ($\eta_{max}$) as a function of $-R_cg_m^{max}$. Black dots are measured at $T=1$\,K by varying $R_c$, while the color dots are obtained via tuning $g_m$ by temperature. 
Data converge to null rectification, thereby validating the back-action at the origin of the supercurrent rectifier. The red dashed line is a reference to the basic model of linear dependence of the critical current on the supercurrent amplitude, which yields $-R_cg_m=d \langle I_c \rangle/dI=\alpha$.}
\label{SDE} 
	\end{figure*}

This scheme is implemented with a nanosized constriction of a superconducting stripe patterned from a niobium (Nb) thin film. Such type of weak link, called Dayem bridge, has shown a reduction of the critical currents up to full suppression under the action of a strong gating effect~\cite{pao19,pug20,pao21}. 
Next to other metallic platforms, such as superconducting nanowires~\cite{sim18,ale21,rit22}, superconductor-normal metal-superconductor Josephson junctions~\cite{sim19,apr21}, and hybrid semiconductor-superconductor weak links~\cite{aka96,doh05,jar06,dam06,xia06,cho13}, Dayem bridges typically display larger critical currents, often reaching fractions of mA, which translate into higher transconductance $g_m$ and, ultimately, in improved rectification efficiencies. 
To date, the proposed mechanisms to account for the phenomenology of gate-controlled supercurrent in metals can be grouped into two main categories:  
i) the supercurrent is suppressed by the nonequilibrium distribution of electronic or phononic excitations which can be induced either by field emission or by leakage currents ~\cite{rit21,rit22,ale21,Gol2021}, ii) the supercurrent suppression arises from the impact of the surface electrostatic potential on the superconducting order parameter~\cite{amo22,sol21,mer20,mer21,roc20,Chak2023}. In this regard, since the back-action mechanism discussed here requires a direct coupling of the gating to the amplitude of the supercurrent, it is valid independently of the mechanism guiding the gating of the supercurrent.

A micrograph of a typical supercurrent rectifier sample is shown in Fig.~\ref{implementation1}c. A constant-thickness strip of niobium (orange) is interrupted by a constriction of length $l\sim 80$ nm and width $w\sim 180$ nm. A side electrode (cyan colored) is placed at a distance $g \sim$ 50\,nm. See Ref.~\cite{mar23} and Methods for basic electrical characterizations and further fabrication details.
Figure~\ref{implementation1}d displays the modulation of the average critical current $\langle I_c\rangle=(I_c^++|I_c^-|)/2$ as a function of the gate bias $V_g^0$ at $T\simeq1$\,K. 
As in former experiments, a plateau is followed by a bipolar decay of the critical current, which is roughly linear in gate voltage $V_g^0$ up to its full suppression at $|V_g^0|\simeq40$\,V.  
The critical current modulation occurs along with a tiny gate-delivered leakage current of a few pA at the onset of the damp of $ I_c $ up to a few nA at full suppression (see Supplementary Figure\,1). 

Figure~\ref{SDE}a-e demonstrates that non-reciprocity of the critical currents emerges for finite gate voltages via a series resistor and can be easily tuned in amplitude. 
We show the switching currents while sweeping the biasing current from zero to positive values ($I_{c}^+$, blue dots) or from zero to negative values ($|I_{c}^-|$, orange dots) as a function of the nominal gate voltage $V_g^0$. The control resistor $R_c$ ranges from 2\,k$\Omega$ 
to 22\,k$\Omega$. 
As $R_c$ increases, the data sets of positive and negative critical currents move apart. 
The nonreciprocal component of the critical current, $\Delta I_{c}= I_{c}^+-|I_{c}^-|$, is plotted in Fig.~\ref{SDE}f-j for the same values of $R_c$. 
The curves are odd-in-$V_g^0$ but nonmonotonic, with peaks at $\simeq \pm 32$\,V that increase in magnitude with $R_c$.
The observed shift and skewness of $I_c^\pm(V_g^0)$ in Fig.\ref{SDE}a-e results from the offset in the gate voltage due to the current flowing through the resistor. Since the gate voltage for complete $I_c$ suppression does not depend, at first order, on $R_c$, yielding an increase or decrease of the slope $\partial I_c / \partial V_g^0$, depending on the polarity of both the critical current and the gate voltage.

To further confirm the role of the control resistor in driving the supercurrent non-reciprocity, we display the critical currents with respect to the gate voltage normalized by the back-action contribution in Fig.~\ref{SDE}k. By defining $V_{g,c}^{\pm} = V_g^0 - I_c^{\pm} R_c$, positive and negative critical currents become identical within the experimental accuracy (see also Supplementary Figure 2).

As previously outlined, the transconductance $g_m$ plays a key role in the performance. Figure~\ref{SDE}l shows the evolution of $g_m =\partial \langle I_c \rangle / \partial V_g^0$ as a function of the gate voltage for $R_c = 2$\,k$\Omega$ and different bath temperatures (see also Supplementary Figure\,3). 
$g_m$ exhibits a peak at $|V_g^0| \simeq 32$\,V and can be tuned by the temperature (see the inset for a maximum of $g_m$ at different temperatures). Its decrease at high $T$ can be ascribed to temperature-induced decay of the critical current~\cite{mar23}.

Finally, it may be convenient to estimate the strength of the back-action. For the toy model developed in the first section, the parameter $\alpha$ weights the dependence of the critical current on the supercurrent and the rectification. The analogous quantity in our experiment is $-R_c g_m$, a dimensionless parameter that quantifies $d  I_c /dI$.
Figure~\ref{SDE}m shows the rectification efficiency evaluated at its maximum $\eta_{max}$ as a function of $-R_c g_m^{max}$. 
A linear trend is observed, as in the model where $\eta = \alpha$ (see Supplementary Figure\,4). Black dots are obtained at 1K by changing the control resistor $R_c$, whereas the data next to the axes origin are measured by tuning the transconductance $g_m$ by changing the bath temperature (color code as in Fig.~\ref{SDE}l). Remarkably, our device reaches a quasi-ideal rectification efficiency, $\eta_{max}\simeq$ 88\,\% for $R_c$= 22\,k$\Omega$ at $T=1$\,K. 
 \begin{figure}[ht]
\centering	
\includegraphics[scale=0.21 ]{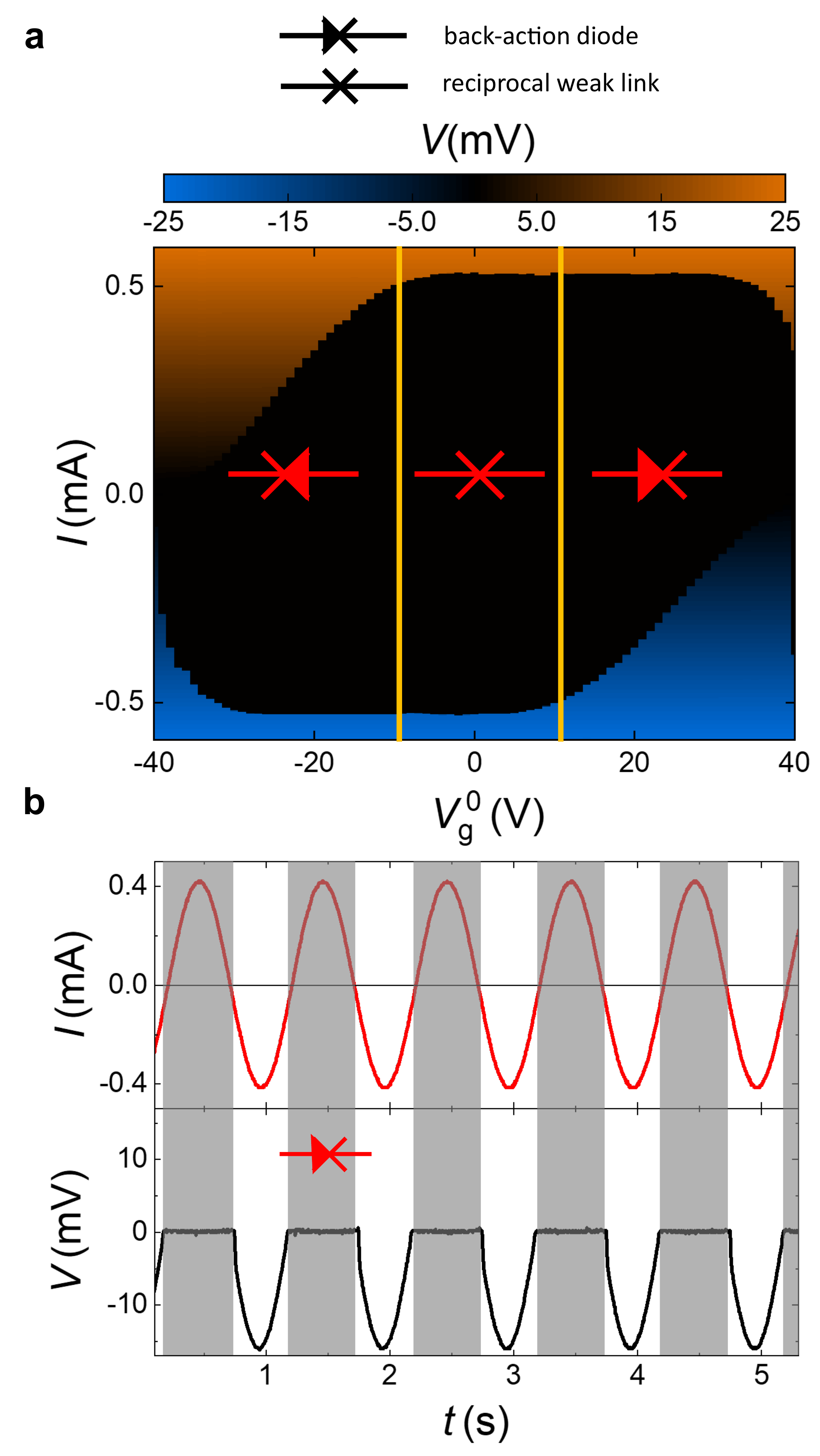}
\caption{
\textbf{Tunability of polarity and half-wave rectification.} \textbf{a,} Contour plot of the voltage drop $V$ versus biasing current $I$ and  gate voltage $V_g^0$ measured at 1K for $R_c=22$\,k$\Omega$. 
Yellow lines mark three regions: the central one, where reciprocal transport takes place and the device behaves as a Josephson reciprocal element (cross symbol), and two external areas beyond the lines, where the polarity-dependent rectification behavior is achieved. At $|V_g^0| \simeq$ 32\,V, very large rectification efficiency ($\eta \simeq 88$\,\%) is achieved. 
\textbf{b}, Half-wave rectifier of the input AC current (upper panel). At $V_g^0=33$\,V and for $R_c$= 12\,k$\Omega$, the output voltage follows the current only in the negative semi-period (i.e., when the weak link is \textit{dissipative}, white bands); on the contrary, $V=0$ in the positive semi-period of $I$ when the Dayem bridge is superconducting (grey bands). White and grey bands slightly differ in width as the rectification is not ideal. The AC biasing current has an
amplitude of 0.4\,mA and frequency of 1\,Hz.}
\label{diode} 
	\end{figure}

We now demonstrate possible applications of our device as a half-wave rectifier with tunable polarity.
Figure~\ref{diode}a presents a color plot of $I-V$ characteristics as a function of the gate voltage $V_g^0$ for a control resistor $R_c=22$\,k$\Omega$. The yellow vertical lines define three working regimes depending on the gate voltage: no sizable rectification, marked by a conventional (reciprocal) Josephson junction symbol; and two non-reciprocal regimes with dominant negative or positive supercurrent, corresponding to the left and right region of the plot respectively.
Therefore, the system operates as a quasi-ideal supercurrent rectifier whose polarity can be changed by simply inverting the sign of the gate voltage.

In the right quadrant of Fig.~\ref{diode}a,  the weak link is predominantly superconducting with a positive bias current but dissipative with a negative bias current. This opens the possibility of exploiting the device as a half-wave rectifier for AC currents.
In principle, this idea holds for supercurrent diodes with any rectification efficiency, till the applied current amplitude lies between $|I_c^-|$ and $I_c^+$. 
However, having a quasi-ideal rectification efficiency, like in our case, widens the range of amplitudes of the injected AC current to be rectified, which greatly relaxes the limitations in applications.
Figure~\ref{diode}b shows the half-wave rectified voltage output $V(t)$ (black trace) and the applied current $I(t)$ of amplitude 0.4\,mA and frequency 1\,Hz (red sinusoidal signal) measured at $V_g^0=33$\,V for $R_c=12$\,k$\Omega$. A zero-voltage output occurs for $I(t)>0$ (superconducting state, gray bands), while the system follows the applied current signal for $I(t)<0$ (dissipative state, white bands).

To show the generality of our work, we have realized the same scheme on a different superconducting platform: an epilayer InAs-based Josephson junction~\cite{pag24}. Here, the gate-controlled supercurrent arises from the charge depletion in the semiconducting weak link. The device is presented in Supplementary Fig.\,5, and exhibits rectification up to 70 $\%$ with an improvement in the power consumption arising from the lower critical current of the Josephson device. More importantly, the system operates without any additional voltage source in the gate electrode due to the asymmetric transconductance with $g_m (V_g^0=0) \neq 0$.

Now, we discuss the technological aspects of the back-action supercurrent rectifiers and their potential impact in the field. 
Possible routes to increase the transconductance $g_m$ and thus the rectification efficiency comprehend the decrease of the gate voltage required for complete supercurrent suppression, the increase of $I_c(V_g^0=0)$, and the modification on the profile of $I_c$ vs $V_g^0$. 
Geometrical designs and material engineering can both be leveraged. Increasing $I_c(V_g^0=0)$ can be accomplished in wider Dayem bridges or semiconductor stacks specifically conceived.
The development of dedicated $I_c$ vs $V_g^0$ profiles or the reduction of the gate voltage for supercurrent suppression looks more challenging in metallic systems. 
The reciprocal pancake-like critical current behavior has resulted in a common feature regardless of the material or sample geometry, and complete suppression, which typically requires tens of volts \cite{ruf2023gate}. 
The latter is also a drawback for implementing gate-controlled superconducting logic circuits ~\cite{lup24} that rely on the transistor output voltage to drive the gate of the following element. 
By contrast, a diode logic family based on back-action supercurrent transistors might lift such constraints by operating at lower gate voltage in the nonreciprocal regime.

Nonreciprocal dissipationless transport has been successfully used as an efficient tool to pinpoint breaking symmetries in novel materials. Our back-action schemes can be exploited to enhance the rectification efficiency that normally exhibits low values for a better understanding of the underlying mechanism or to explore polarity-dependent phenomena. However, the field of superconducting diodes has rapidly progressed in developing platforms aimed at enhancing efficiency and applications for quantum technology~\cite{upa24}. Currently, scalability and integration, major limiting factors, have been addressed in vortex-based~\cite{cas24} and ferromagnetic~\cite{ing24} diodes shunted by dissipative elements. In our approach, the resistor can be placed at any stage of the cryostat, effectively minimizing the heat load. In this context, the superior performance and magnetic-free back-action rectifier offer a significant advance in the field compatible with extremely field-sensitive qubits or in noisy magnetic environments~\cite{pao21}.\\ 

\noindent\large{\textbf{Conclusions}}\normalsize\\
In summary, we have proposed and implemented a supercurrent back-action rectifier. To date, most efforts devoted to lifting the odd parity in the current-phase relation in superconducting weak links rely on magnetic exchange interactions ~\cite{rya01,sic12} and magnetic fields in combination with sizable spin-orbit coupling~\cite{ke19,dam06,ass19,str20}, or nonequilibrium conditions~\cite{mar23a}.
Instead, the original class of devices presented here counts on the self-induced effect of the supercurrent amplitude to break the parity condition.
Though intrinsically reciprocal, supercurrent transport across the metallic weak link is made nonreciprocal by the retroaction of the bias current on the weak link itself. 
This results in tuning the critical current by a gate voltage sensitive to the current polarity.
The rectification efficiency is easily adjustable and can be boosted nearly to one.
Such reduction of fabrication complexity, combined with the absence of magnetic fields, represents a clear asset for scalability and integration with other superconducting devices, making our platform attractive for next-generation superconducting electronics.\\
The generality of the scheme is also demonstrated in a hybrid semiconducting-superconducting field-effect transistor. The substantial decrease in power consumption and the absence of a gate voltage source are real assets for Josephson-based back-action supercurrent rectifiers as a building block in future nonreciprocal superconducting technologies.

We expect similar devices to be implemented on all the superconducting platforms adjustable by other control knobs, such as
phonon-controlled Josephson junctions~\cite{hutchinson2004controlled,pod07}, flux-retroacted 
superconducting quantum interference devices~\cite{handbook}, or quasiparticle injection-driven superconducting transistors~\cite{ber14,giazotto2005josephson,tirelli2008manipulation,morpurgo1998hot,baselmans1999reversing}.
Finally, other functionalities are expected beyond the linear back-action approximation used in this work for rectification improvements or multi-valued circuits.\\

\noindent\large{\textbf{Methods}}\normalsize\\
\textbf{Sample fabrication.}
Nb strips and constrictions are patterned by e-beam lithography on AR-P 679.04 (PMMA) resist. PMMA residuals are removed by O$_2$-plasma etching after developing. Nb thin films of 25\,nm thickness were deposited by sputtering at a base pressure of 2 $\times$ 10$^{-8}$ Torr in a 4 mTorr Ar (6N purity) atmosphere and liftoff by AR-P 600.71 remover. Details of the fabrication of InAs-based Josephson junction can be found elsewhere~\cite{pag24}.\\ 

\noindent\textbf{Transport measurements.}
Transport measurements were carried out in filtered (two-stage RC and $\pi$ filters) cryogen-free $^3$He-$^4$He dilution refrigerators by a standard 4-wire technique. DC current-voltage characteristics were measured by sweeping a low-noise current bias positively and negatively, and by measuring the voltage drop occurring across the weak links with a room-temperature low-noise pre-amplifier. 
The switching currents and error bars were obtained from 10-20 reiterations of the $IV$ curves,
and their accuracy is mostly given by the current step set to $\Delta I < 0.002 I_c(V_g^0=0)$, where $I_c$ is the switching current of the Nb nanobridge. 
Joule heating in the system is minimized by automatically switching the current off once the device turns into the normal state. 
A delay between sweeps was optimized to keep the stability of the fridge temperature lower than 50 mK. 
Furthermore, no changes in the switching currents (up to the accuracy given by the standard deviation) were observed in different cooling cycles, by changing the order of the sweeps, or by adding an extra delay in the acquisition protocol, thereby concluding that hysteretic behavior or local heating is negligible. 
The gate-nanobridge current $I_g$ was acquired by using a low-noise voltage source and a $10^{-11}$\,A/V-gain low-noise current amplifier in a two-wire configuration.\\

\noindent\large{\textbf{Data availability}}\normalsize\\
\noindent \footnotesize{The data that support the findings of this study are available
from the corresponding author upon reasonable request.}\\

\noindent\large{\textbf{References}}\normalsize
\vspace{-1.5cm}
%
\vspace{1cm}
\noindent\large{\textbf{Acknowledgments}}\footnotesize\\
This work was funded by the EU’s Horizon 2020 Research and Innovation Framework Program under Grant
Agreement No. 964398 (SUPERGATE), No. 101057977 (SPECTRUM), by the PNRR MUR project PE0000023-NQSTI, and by the NextGenerationEU PRIN project 2022A8CJP3 (GAMESQUAD).\\

\noindent\large{\textbf{Author contributions}}\footnotesize\\
D.M. fabricated the samples, conducted the experiments, and analyzed data with inputs from A.C., E.S., and F.G. The theoretical models were developed by Y.F., M.T.M., E.S., M.O., and M.C. 
L.B., A.P., G.S, L.S, fabricated and characterized InAs-based superconducting rectifier. The manuscript was written by
D.M., A.C., and M.C., with inputs from all the authors. D.M., E.S., and F.G. conceived the experiment. F.G. supervised and coordinated the project. 
All authors discussed the results and their implications equally at all stages.\\

\noindent\large{\textbf{Competing interests}}\footnotesize\\
The authors declare no competing interests.\\

\noindent\large{\textbf{Additional Information}}\footnotesize\\
\textbf{Supplementary Information} The online version contains supplementary material. \\

\noindent\textbf{Correspondence} and requests for materials should be addressed to Daniel Margineda, Elia Strambini, and Francesco Giazotto.

\onecolumngrid
\newpage
\section{Supplementary Information of ``Back-action supercurrent rectifiers"}

\noindent\large{\textbf{Supplementary Note }}\normalsize\\
\textbf{Current-phae relation symmetries due to supercurrent back-action.}\\
In this section, we will demonstrate that the back-action mechanism on the supercurrent i) will not produce any spontaneous phase, i.e, $I(\phi=0)=0$ and ii) rectification of the supercurrent will emerge for any functional form of the effective Josephson coupling $E_J(I)$ with an odd parity.
To this aim, we start considering the free energy of a Josephson junction that can be cast into the form of

\begin{equation*}
F_{J}=E_{J} [1-\cos(\phi)]
\end{equation*}
with $E_J$ the Josephson coupling and $\phi$ the phase difference among the superconductors forming the junction.
For conventional configurations, $E_J$ depends on factors related to the junction (e.g. material, barrier, etc). From the expression of the free energy, one can directly deduce the supercurrent flowing across the junction as the variation of the free energy with respect to the phase bias (in units of $\frac{2 e}{\hbar}$):
\begin{equation}
I(\phi)=\frac{\partial F_J}{\partial \phi}
\label{cpr}
\end{equation}
that yields the well-known Josephson relation $I(\phi)=I_c \sin(\phi)$ with $I_c=E_{J}$ being the critical current, namely, the maximal current that can sustain the junction without any voltage drop. This relation sets out the dc Josephson effect with a supercurrent existing between two superconductors that are coupled through a thin layer.

Now, we introduce the back-action mechanism on the amplitude of the supercurrent. We assume that the coupling among the superconductors depends on the amplitude of the supercurrent flowing across the junction. Hence, one can postulate a free energy of the type:
\begin{equation}
F_J= E_{J}(I) [1-\cos(\phi)]
\label{free_en_1}
\end{equation}
where the coupling $E_{J}(I)$ is a function of the supercurrent $I$ flowing through the superconductors.
Before discussing the structure of the coupling, we would like to emphasize that Eq.~\ref{free_en_1} describes a Josephson system with a transmission probability of Cooper pairs that depends on the amplitude of the supercurrent flowing across the junction. The form of $F_J$ is consistent with the physical requirement that at zero applied phase bias ($\phi=0)$ the free energy is constant (i.e. $F_J(\phi=0)=0$) at any value of the supercurrent flowing through the junction. As a consequence, the self-induced supercurrent model fulfills the relation $I(0)=0$, i.e. there is no spontaneous supercurrent flowing at zero phase bias. Such result can be deduced from the supercurrent expression obtained from  Eq.~\ref{cpr} and Eq.~\ref{free_en_1}:
\begin{equation}
I(\phi)=\frac{\partial E_{J}(I)}{\partial I} \frac{\partial I(\phi)}{\partial \phi} [1-\cos(\phi)]+E_{J}(I) \sin{\phi} 
\label{self}
\end{equation}
There, by making the limit for $\phi\rightarrow 0$ on the two sides of the Eq.~\ref{self}, the $I(0)=0$ constraint is immediately deduced independently of the functional form for $E_{J}(I)$.

The second important feature refers to the connection between the back-action mechanism and the possibility of achieving a rectification of the supercurrent, i.e., the maximal positive forward amplitude turns out to be different from the maximal negative backward amplitude of the supercurrent.

For this issue, we consider the dependence of the Josephson coupling on the supercurrent amplitude in two different cases: i) $E_{J}(I)$ is an even parity function with respect to $I$, $E_{J}(-I)=E_{J}(I)$. ii) $E_{J}(I)$ is an odd parity function with respect to $I$, $E_{J}(-I)=-E_{J}(I)$.

In the first scenario, we can observe that for an effective coupling  $E_{J}(I)$ with even parity, the solution for $I(\phi)$ will present a definite and odd parity with respect to $\phi$, i.e. $I(\phi)=-I(-\phi)$.  This conclusion can be deduced by analyzing the parity of Eq.~\ref{self} and constructing the solution in an iterative way with respect to the strength of the back-action amplitude coupling. 
Hence, for an effective coupling that has an even parity dependence in the supercurrent amplitude, the current phase relation yields a reciprocal transport. 

In the second scenario ii), with $E_{J}(I)$ being an odd parity function with respect to $I$, the solution of the Eq. \ref{self} will give a current phase relation that does not have a definite parity in the phase bias $\phi$ and therefore nonreciprocal transport.
We can arrive to this conclusion without solving the self-consistent equations from the first derivative of the supercurrent
\begin{eqnarray}
\frac{\partial I(\phi)}{\partial \phi}=\frac{I(\phi)-E_{J}(I) \sin(\phi)}{(1-\cos\phi) \frac{\partial E_{J}(I)}{\partial I}} 
\label{derivative}\,. 
\end{eqnarray}
obtained from Eq.~\ref{self}.

Maximum amplitude for the positive $I_c^+$ and negative $I_c^-$ supercurrents is given by the condition $\frac{\partial I(\phi)}{\partial \phi}=0$ 
(assuming that there are no discontinuities in the current phase relation).
 Since $I(\phi)$ does not have a definite parity in the variable $\phi$, $I_c^+=I(\phi_+)\neq I_c^-=I(\phi_-)$.
This result implies that the supercurrent will be nonreciprocal for any functional form of the effective coupling $E_{J}(I)$ that includes odd parity terms with respect to the supercurrent amplitude $I$.

The model can be also extended to non-tunnel superconducting circuits in which the free energy may be more complex with respect to the simple cosine function. In particular, for reciprocal junctions, the cosine function will be substituted by a generic periodic even function, and through simple parity constraints the same conclusions can be extracted. 
\\
\newpage

\setcounter{figure}{0}
\renewcommand{\figurename}{\textbf{
Supplementary Figure}}
\noindent\large{\textbf{Supplementary Figures }}\normalsize\\

\begin{figure*}[ht]
\includegraphics[scale=0.28]{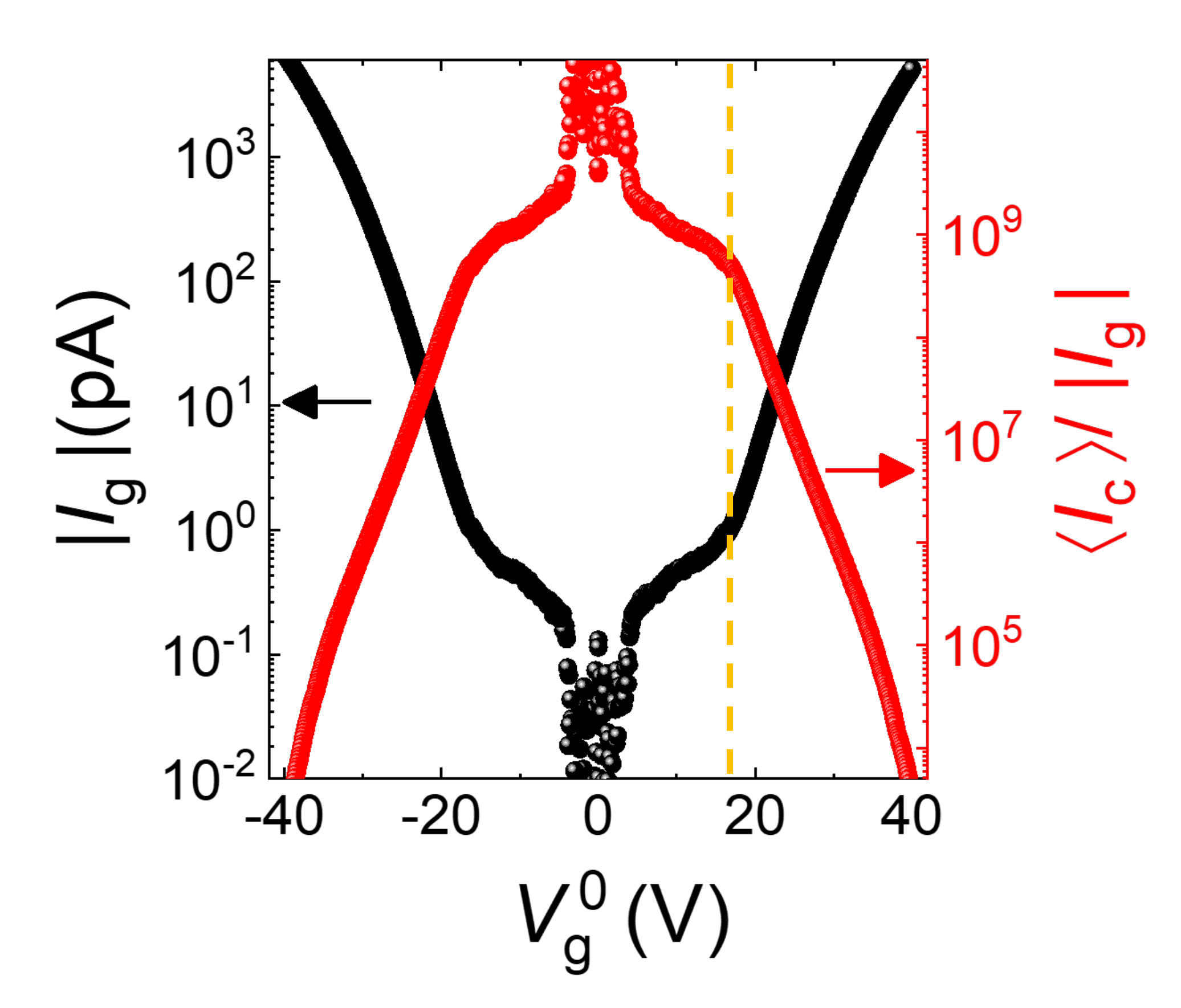}
\caption{\textbf{Gate current.} 
Gate-delivered leakage current to the weak link $I_g$ at T =1\,K (black dots). $I_g$ increases from a few pA at the onset of the critical current $\langle I_c\rangle$ suppression (yellow dashed line) to a few nA at full suppression. $\langle I_c\rangle/I_g$ ratio is given by red dots.}
\label{SI1} 
\end{figure*}

\begin{figure*}[ht]
\includegraphics[scale=0.25]{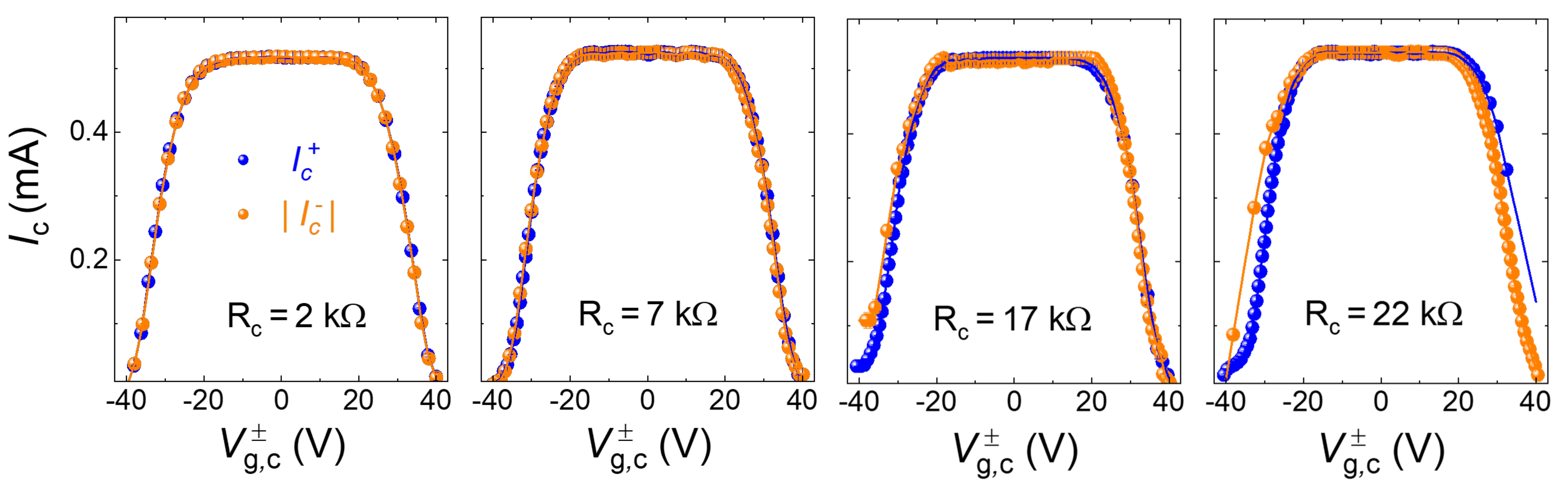}
\caption{\textbf{Back-action normalization.}  Quasi-symmetric behavior of the critical currents obtained by normalizing the gate voltage with the back-action contribution, $V_{g,c}^{\pm}=V_g -I_{c}^{\pm}R_c$ for different control resistors $R_c$.}
\label{SI2} 
\end{figure*}

\begin{figure*}[ht]
\includegraphics[scale=0.28]{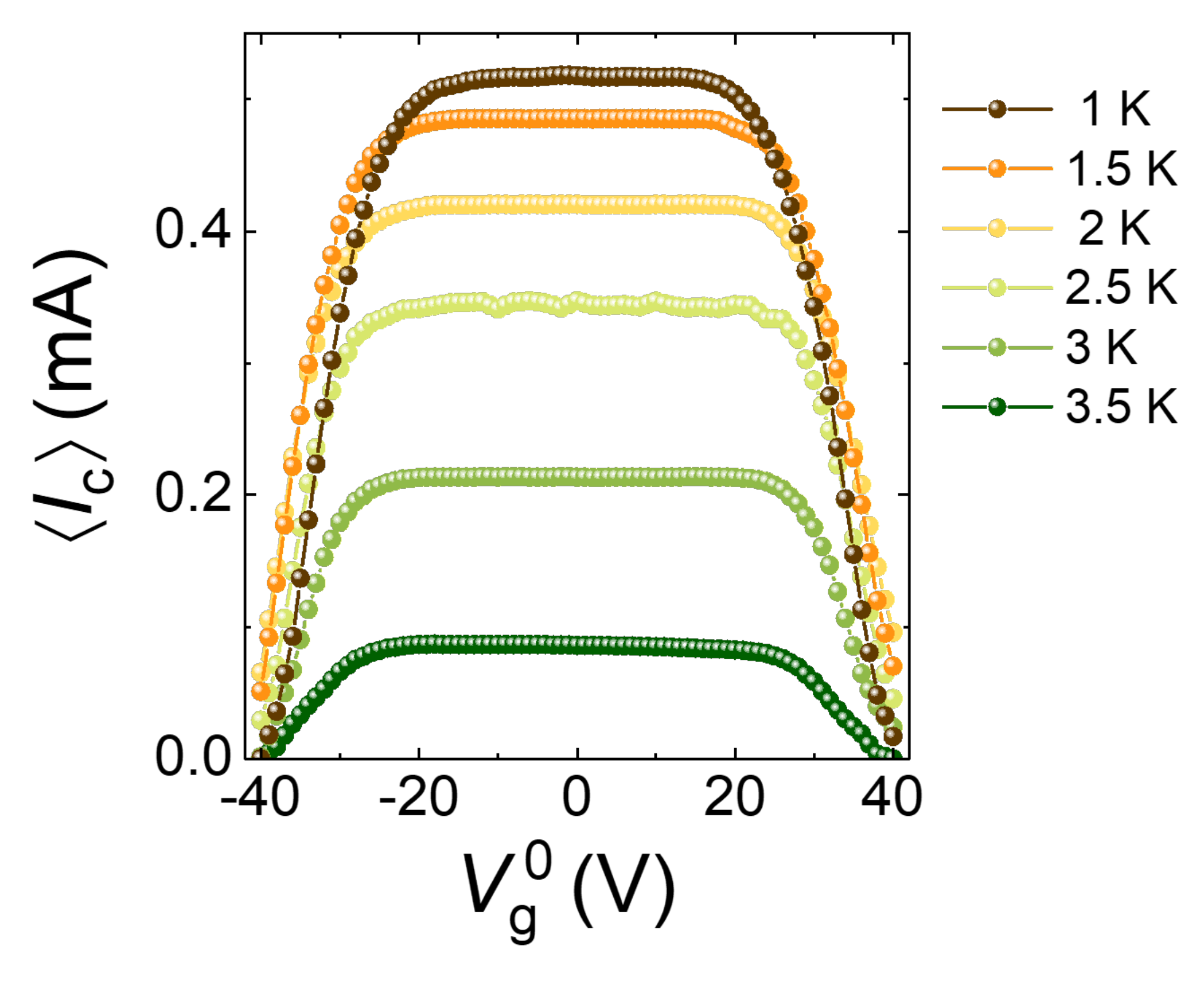}
\caption{\textbf{Temperature dependence of the gate-controlled supercurrent.} Average critical current $\langle I_c\rangle$ as a function of the gate voltage $V_g^0$ for selected values of bath temperature.}
\label{SI3} 
\end{figure*}

\begin{figure*}[ht]
\includegraphics[scale=0.16 ]{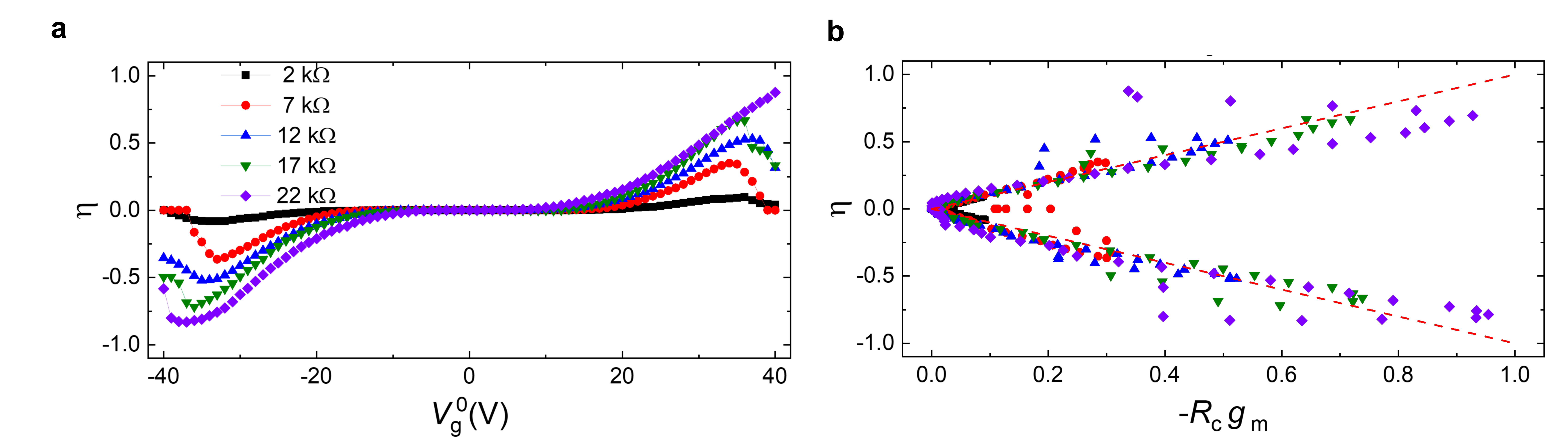}
\caption{\textbf{Diode rectification efficiency.}
\textbf{a,} Diode rectification parameter $\eta$ as a function of the applied gate voltage $V_g^0$ for different control resistors $R_c$. \textbf{b,} $\eta$ as a function of the back-action parameter $\alpha=-R_cg_m$ extracted from \textbf{a,} with the same color code. Negative rectification efficiency corresponds to negative gate voltages.  Red dashed lines represent the ideal rectification $\eta=\alpha$ given by the linear model.}
\label{SI} 
\end{figure*}

\begin{figure*}
\includegraphics[scale=0.15 ]{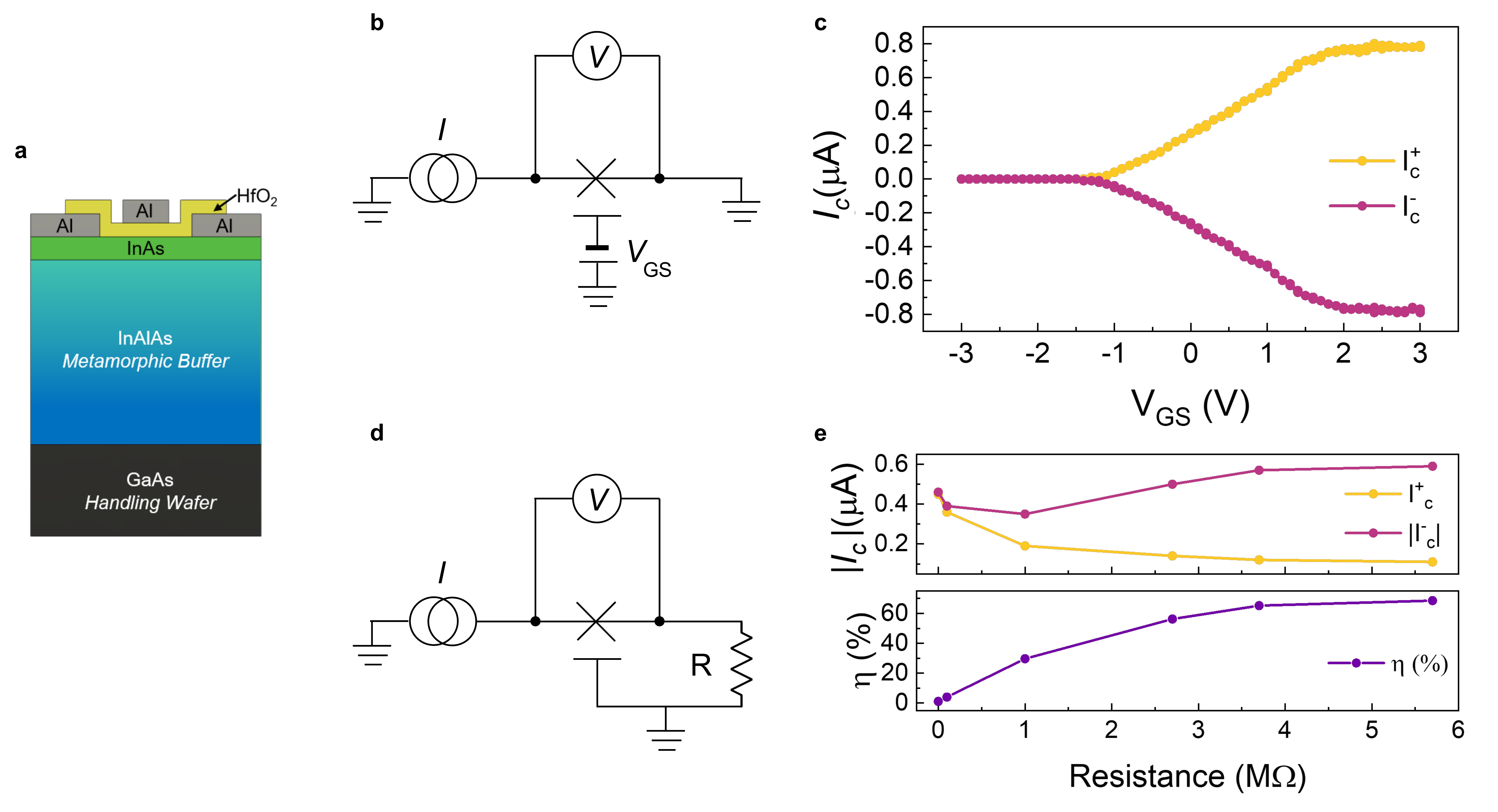}
\caption{
\textbf{Back-action supercurrent diode implemented in an InAsOI Josephson junction.} 
\textbf{a} Stack of the InAsOI heterostructure. The InAs epilayer is grown on an InAlAs metamorphic buffer layer which acts as a cryogenic insulator and avoids lattice
mismatch between the InAs and the GaAs handling substrate.  The InAs epilayer supports supercurrent via the
proximity effect inherited by the Al superconducting leads. Further details are reported in A. Paghi, et al., arXiv 2405.07630. The oxide (HfO$_2$) isolates the Al top gate, which controls the charge in the InAs channel and the resulting critical current. 
\textbf{b} Schematic of the electrical circuitry to measure the gate-dependent I-V characteristics of the junction. \textbf{c} Critical currents with positive and negative current bias sweep ($I_C^+$ and $I_C^-$, respectively) as a function of the gate voltage $V_{GS}$. \textbf{d} Schematic of the circuitry to realize a back-action supercurrent diode. No voltage source is necessary to the gate since $I_c$ vs $V_{GS}$ is not even in $V_{GS}$, see panel c. \textbf{e} Top panel: critical currents with positive and negative current bias sweep measured at 100~mK for different values of the control resistance, showing the diode effect. Bottom panel: supercurrent diode efficiency.}
\label{InAsDIODE} 
\end{figure*}

\end{document}